\documentclass[12pt]{article}

\begin{document}
\vspace*{-.6in}
\thispagestyle{empty}
\begin{flushright}
CALT-68-2091\\
hep-th/9701008
\end{flushright}
\baselineskip = 20pt

\vspace{.5in}
{\Large
\begin{center}
Coupling a Self-Dual Tensor to Gravity in Six Dimensions\footnote{Work
supported in part by the U.S. Dept. of Energy under Grant No.
DE-FG03-92-ER40701.}
\end{center}}

\begin{center}
John H. Schwarz\\
\emph{California Institute of Technology, Pasadena, CA  91125, USA}
\end{center}
\vspace{1in}

\begin{center}
\textbf{Abstract}
\end{center}
\begin{quotation}
\noindent  A recent result concerning interacting theories of self-dual
tensor gauge fields in six dimensions is generalized to include coupling to
gravity.  The formalism makes five of the six general coordinate invariances
manifest, whereas the sixth one requires a non-trivial analysis.  The result
should be helpful in formulating the world-volume action of the M theory
five-brane.
\end{quotation}
\vfil

\newpage

\pagenumbering{arabic}

In a recent paper, Perry and the author studied interacting theories of a
self-dual tensor gauge field in six dimensions~\cite{perry}.  Since there is no
straightforward way to describe such a theory with manifest Lorentz invariance,
we chose to use a formalism in which Lorentz invariance was manifest only in a
5d subspace, while the remaining Lorentz symmetries were realized in a more
subtle non-manifest way.  The purpose of the present paper is to extend that
work to include the coupling to a 6d metric field.  The resulting theories will
have manifest general coordinate invariance in five directions and non-manifest
general coordinate invariance in the sixth one.

There are two different motivations and potential applications of these
results.  The first is to 6d supergravity theories.  The minimal supergravity
multiplet contains the self-dual tensor gauge field in addition to the graviton
and gravitino fields.  The second application is to the construction of the
world-volume action of the M theory five-brane, since this is also a 6d theory with a
self-dual tensor gauge field.  The metric in this case is induced from the
superspace coordinates of the ambient 11d space-time.

In the following we denote 6d coordinates by $x^{\hat\mu} = (x^\mu, x^5)$,
where $\mu = 0,1,2,3,4$.  The $x^5$ direction is singled out as the one that
will be treated differently from the other five.  The 6d metric
$G_{\hat\mu\hat\nu}$ contains 5d pieces $G_{\mu\nu}, G_{\mu 5}$, and $G_{55}$.
All formulas will be written with manifest 5d general coordinate invariance.
As in Ref.~\cite{perry}, we represent the self-dual tensor gauge field by a
$5\times 5$ antisymmetric tensor $B_{\mu\nu}$.  The way this comes about can be
described quite easily for the free theory.  If one starts with a 6d tensor
$B_{\hat\mu\hat\nu}$, with field strength $H_{\hat\mu \hat\nu\hat\rho} = 3
\partial_{[\hat\mu} B_{\hat\nu \hat\rho]}$, then the self-duality condition is
the first-order field equation
\begin{equation}
H_{\hat\mu \hat\nu \hat\rho} = {1\over 6\sqrt{-G}} G_{\hat\mu \hat\mu'}
G_{\hat\nu \hat\nu'} G_{\hat\rho \hat\rho'} \epsilon^{\hat\mu' \hat\nu'
\hat\rho' \hat\lambda \hat\sigma \hat\eta} H_{\hat\lambda \hat\sigma \hat\eta},
\end{equation}
where $G = {\rm det}\, G_{\hat\mu \hat\nu}$.  It is possible to solve this equation for
$H_{\mu\nu 5}$ in terms of $G_{\hat\mu \hat\nu}$ and $H_{\mu\nu\rho}$:
\begin{equation}
H_{\mu\nu 5} = K_{\mu\nu} (G, H).
\end{equation}
Since
\begin{equation}
H_{\mu\nu 5} = \partial_5 B_{\mu\nu} + \partial_\mu B_{\nu 5} - \partial_\nu
B_{\mu 5},
\end{equation}
taking a curl eliminates $B_{\mu 5}$ leaving a 
second-order field equation that involves
$B_{\mu\nu}$ only:
\begin{equation}
{1\over 2} \epsilon^{\mu\nu\rho\lambda\sigma} \partial_\rho K_{\lambda\sigma} =
{1\over 2} \epsilon^{\mu\nu\rho\lambda\sigma} \partial_5 \partial_\rho
B_{\lambda\sigma} = \partial_5 \tilde{H}^{\mu\nu}, \label{fieldeqn}
\end{equation}
where
\begin{equation}
\tilde{H}^{\mu\nu} = {1\over 6} \epsilon^{\mu\nu\rho\lambda\sigma}
H_{\rho\lambda\sigma}.
\end{equation}

It is a straightforward exercise to formulate a Lagrangian that gives the
field equation~(\ref{fieldeqn}).  The result is $L = L_1 + L_2 + L_3$, where 
\begin{eqnarray}
L_1 &=& {\sqrt{-G} \over 2(-G_5)}{\rm tr}(G\tilde{H} G\tilde{H}), \nonumber \\
L_2 &=& {1\over 2} \tilde{H}^{\mu\nu} \partial_5 B_{\mu\nu}, \\
L_3 &=& - {1\over 4}
\epsilon_{\mu\nu\rho\lambda\sigma} {G^{5\rho}\over G^{55}} \tilde{H}^{\mu\nu}
\tilde{H}^{\lambda\sigma}.\nonumber 
\end{eqnarray}
This result was given
previously in Ref.~\cite{schwarz}.  The only difference is that there the time
direction was treated as special, whereas now we are singling out a spatial
direction.
Note that $L_2$ contains the only $x^5$ derivative, and it does not depend on
the metric.  The notation is as follows:  $G_5$ is the 5d determinant $(G_5 =
{\rm det}\, G_{\mu\nu})$, while $G^{55}$ and $G^{5\rho}$ are components of the inverse
6d metric $G^{\hat\mu \hat\nu}$.  The trace only involves 5d indices:
\begin{equation}
{\rm tr} (G\tilde{H} G\tilde{H}) = G_{\mu\nu} \tilde{H}^{\nu\rho} G_{\rho\lambda}
\tilde{H}^{\lambda\mu}.
\end{equation}
The $\epsilon$ symbols are purely numerical with $\epsilon^{01234} = 1$ and
$\epsilon^{\mu\nu\rho\lambda\sigma} = - \epsilon_{\mu\nu\rho\lambda\sigma}$.  A
useful relation is $G_5 = G G^{55}$.

Our goal now is to generalize the preceding free theory result to the case
of an interacting theory.  For this purpose we consider the generalization
\begin{equation}
L_1 = \sqrt{-G} f(z_1, z_2) 
\end{equation}
while keeping $L_2$ and $L_3$ unchanged.  The $z$ variables are defined to be
\begin{eqnarray}
z_1 &=& {{\rm tr} (G\tilde{H} G\tilde{H})\over 2( -G_5)}\nonumber \\
z_2 &=& {{\rm tr} (G\tilde{H} G\tilde{H} G\tilde{H} G\tilde{H})\over 4 (-G_5)^2}.
\end{eqnarray}
This is quite general, since traces of higher powers of
$G\tilde{H}$ can be reexpressed in
terms of $z_1$ and $z_2$.  The quantities $z_1$ and $z_2$
are scalars under 5d general coordinate
transformations.  Note that the free theory result corresponds to $f = z_1$.
Our goal now is to determine the class of functions $f$ for which the theory
has 6d general coordinate invariance.

Infinitesimal parameters of general coordinate transformations are denoted
$\xi^{\hat\mu} = (\xi^\mu, \xi)$.  Since 5d general coordinate invariance is
manifest, we focus on the $\xi$ transformations only.  The metric transforms in
the standard way
\begin{equation}
\delta_\xi G_{\hat\mu \hat\nu} = \xi \partial_5 G_{\hat\mu \hat\nu} +
\partial_{\hat\mu} \xi G_{5\hat\nu} + \partial_{\hat\nu} \xi G_{\hat\mu 5}.
\label{Gvar}
\end{equation}
However, something non-standard is required for $B_{\mu\nu}$.  If we had an
ordinary 6d tensor $B_{\hat\mu \hat\nu}$, the standard transformation would be
$\delta B_{\hat\mu \hat\nu} = \xi^{\hat\rho} H_{\hat\rho \hat\mu
\hat\nu}$, which gives, in particular, $\delta_\xi B_{\mu\nu} = \xi
H_{5\mu\nu}$.  But $H_{5\mu\nu}$ contains fields that are not part of our
theory.  To see what is going on, let us form the $B_{\mu\nu}$ field equation.
It is given again by eq. (\ref{fieldeqn}), where now
\begin{equation}
K_{\mu\nu} = -{\partial (L_1 + L_3) \over\partial \tilde{H}^{\mu\nu}} =
K_{\mu\nu}^{(1)} f_1+ K_{\mu\nu}^{(2)} f_2+ K_{\mu\nu}^{(\epsilon)}
\end{equation}
with
\begin{eqnarray}
K_{\mu\nu}^{(1)} &=& {\sqrt{-G} \over (-G_5)}{(G\tilde{H} G)_{\mu\nu}}
\nonumber \\
K_{\mu\nu}^{(2)} &=& {\sqrt{-G} \over (-G_5)^2}{(G\tilde{H} G\tilde{H} G\tilde{H}
G)_{\mu\nu}}  \\
K_{\mu\nu}^{(\epsilon)} &=& \epsilon_{\mu\nu\rho\lambda\sigma}
{G^{5\rho}\over 2 G^{55}} \tilde{H}^{\lambda\sigma}, \nonumber
\end{eqnarray}
and we have defined
\begin{equation}
f_i = {\partial f\over\partial z_i} , \quad i = 1,2.
\end{equation}
There is now a natural guess for $\delta_\xi B_{\mu\nu}$, since the field
equation implies that $H_{5\mu\nu}$ is equivalent to $K_{\mu\nu}$ on shell.
Therefore we postulate the transformation law
\begin{equation}
\delta_\xi B_{\mu\nu} = \xi K_{\mu\nu}. \label{Bvar}
\end{equation}

The next step is to examine the symmetry
by computing the variation of the action under the
transformations $\delta_\xi G_{\hat\mu \hat\nu}$ and $\delta_\xi B_{\mu\nu}$
given in eqs.~(\ref{Gvar}) and (\ref{Bvar}).  
We will speak of the variation of $L$, but since it is the action
we are really interested in, total derivatives will be dropped whenever
convenient.  The variation is a sum of three pieces
\begin{equation}
\delta L = \delta_G (L_1  +L_3) + \delta_B (L_1 + L_3) + \delta_B L_2.
\end{equation}
The $B$ variations are relatively simple:
\begin{eqnarray}
\delta_B (L_1 + L_3) &=& (\delta_\xi \tilde{H}^{\mu\nu}) {\partial
(L_1 + L_3) \over\partial\tilde{H}^{\mu\nu}} = -{1 \over 2}
 \epsilon^{\mu\nu\rho\lambda\sigma}
\partial_\rho (\xi K_{\lambda\sigma}) K_{\mu\nu} 
\sim -{1\over 4} \epsilon^{\mu\nu\rho\lambda\sigma} K_{\mu\nu}
K_{\lambda\sigma} \partial_\rho \xi, \nonumber \\
\delta_B L_2 &\sim & - \delta_\xi B_{\mu\nu} \partial_5 \tilde{H}^{\mu\nu} = - \xi
K_{\mu\nu} \partial_5 \tilde{H}^{\mu\nu}.
\end{eqnarray}
The symbol $\sim$ means that a total derivative has been dropped.
Somewhat more complicated is
\begin{eqnarray}
\delta_G (L_1 + L_3) &=& \partial_5 (\xi \sqrt{-G}) f + 
\sqrt{-G} \delta_G f
- {1\over 4} \epsilon_{\mu\nu\rho\lambda\sigma} \delta \left({G^{5\rho}\over
G^{55}}\right) \tilde{H}^{\mu\nu} \tilde{H}^{\lambda\sigma} \nonumber \\
&\sim& \sqrt{-G} \Big(-\xi (f_1 \partial_5 z_1 + f_2 \partial_5 z_2) + f_1 \delta_G
z_1 + f_2 \delta_G z_2\Big) \nonumber \\
& &- {1\over 4} \epsilon_{\mu\nu\rho\lambda\sigma} \left[\partial_5 \left(\xi
{G^{5\rho}\over G^{55}}\right)  + \partial_\eta \xi \left({2G^{5\eta}
G^{5\rho}\over (G^{55})^2} - {G^{\eta\rho}\over G^{55}}\right) \right]
\tilde{H}^{\mu\nu} \tilde{H}^{\lambda\sigma}. \label{Gvariation}
\end{eqnarray}
Here one must insert
\begin{equation}
\delta_G z_1 = \xi \partial_5 z_1 + \xi (-G_5)^{-1} \partial_5
\tilde{H}^{\mu\nu} (G\tilde{H} G)_{\mu\nu} 
+ 2 \partial_\mu \xi \left({G^{5\mu}\over G^{55}} z_1 + {G_{5\nu}\over
(-G_5)} (\tilde{H} G \tilde{H})^{\mu\nu}\right)
\end{equation}
and
\begin{eqnarray}
\delta_G z_2 &=& \xi \partial_5 z_2 + \xi (-G_5)^{-2} \partial_5
\tilde{H}^{\mu\nu} (G \tilde{H} G \tilde{H} G \tilde{H} G)_{\mu\nu} \nonumber\\
&{}& +2\partial_\mu \xi \left( {2G^{5\mu}\over G^{55}} z_2 + {G_{5\nu}\over
(-G_5)^2} (\tilde{H} G \tilde{H} G \tilde{H} G \tilde{H})^{\mu\nu}\right).
\end{eqnarray}

Now we collect terms and simplify.  For example, the terms proportional to
$f_1$ are
\begin{eqnarray}
&-& \xi K_{\mu\nu}^{(1)} f_1\partial_5 \tilde{H}^{\mu\nu} - {1\over 2}
\epsilon^{\mu\nu\rho\lambda\sigma} \partial_\rho \xi K_{\mu\nu}^{(1)} f_1
K_{\lambda\sigma}^{(\epsilon)}
+ \xi {\sqrt{-G}\over (-G_5)} \partial_5 \tilde{H}^{\mu\nu} (G \tilde{H}
G)_{\mu\nu} f_1\nonumber \\
&+& 2 \partial_\mu \xi \sqrt{-G} \left({G^{5\mu}\over G^{55}} z_1 +
{G_{5\nu}\over (-G_5)} (\tilde{H} G \tilde{H})^{\mu\nu}\right)f_1. \label{f1terms}
\end{eqnarray}
The first and third terms, which are proportional to $\xi$, 
cancel.  To compare the other two terms we
use
\begin{equation}
\epsilon^{\mu\nu\rho\lambda\sigma} K_{\lambda\sigma}^{(\epsilon)} = -
\delta_{\mu' \nu' \rho'}^{\mu\nu\rho} {G^{5 \rho'}\over G^{55}}
\tilde{H}^{\mu'\nu'}.
\end{equation}
Then, using $G^{5\mu} G_{\mu\nu} = - G^{55} G_{5\nu}$, it follows that the
second and fourth terms also cancel.  The terms proportional to $f_2$ cancel in the
same way.

The remaining terms proportional to $\xi$ are 
$-\xi K_{\mu\nu}^{(\epsilon)} \partial_5 \tilde{H}^{\mu\nu}$
and the term that results from partially integrating the $\partial_5$ term
in the last line of eq.~(\ref{Gvariation}). These two terms cancel.

The terms that now remain are all proportional to $\partial_\mu \xi$:
\[
{1\over 4} \epsilon_{\mu\nu\rho\lambda\sigma} \partial_\eta \xi
\left({G^{\eta\rho}\over G^{55}} -2 {G^{5\eta} G^{5\rho}\over
(G^{55})^2}\right) \tilde{H}^{\mu\nu} \tilde{H}^{\lambda\sigma} - {1\over 4}
\epsilon^{\mu\nu\rho\lambda\sigma} \partial_\rho \xi K_{\mu\nu}^{(\epsilon)}
K_{\lambda\sigma}^{(\epsilon)}\]
\begin{equation}
- {1\over 4} \epsilon^{\mu\nu\rho\lambda\sigma} \partial_\rho \xi
(K_{\mu\nu}^{(1)} f_1 + K_{\mu\nu}^{(2)} f_2) (K_{\lambda\sigma}^{(1)} f_1 +
K_{\lambda\sigma}^{(2)} f_2). \label{remains}
\end{equation}
The two terms on the first line of this expression combine to give
\begin{equation}
W = \epsilon_{\mu\nu\rho\lambda\sigma} \partial_\eta \xi {G_5^{\eta\rho}\over
4G^{55}} \tilde{H}^{\mu\nu} \tilde{H}^{\lambda\sigma},
\end{equation}
where
\begin{equation}
G_5^{\eta\rho} = G^{\eta\rho} - {G^{5\eta} G^{5\rho}\over G^{55}}
\end{equation}
is the inverse of the $5 \times 5$ matrix $G_{\eta\rho}$.

It remains to evaluate the terms on the second line of (\ref{remains}).  Let us begin with
the term proportional to $f_1^2$.  Using the identity
\begin{equation}
\epsilon^{\mu\nu\rho\lambda\sigma} G_{\mu\mu'} G_{\nu\nu'} G_{\lambda \lambda'}
G_{\sigma\sigma'} = - G_5 \epsilon_{\mu' \nu' \rho' \lambda' \sigma'}
G_5^{\rho\rho'},
\end{equation}
one finds that
\begin{equation}
- {1\over 4} \epsilon^{\mu\nu\rho\lambda\sigma} \partial_\rho \xi
K_{\mu\nu}^{(1)} K_{\lambda\sigma}^{(1)} f_1^2 = - f_1^2 W.
\end{equation}
Using the same identity in the $f_1 f_2$ term gives
\begin{equation}
- {1\over 2} \epsilon^{\mu\nu\rho\lambda\sigma} \partial_\rho \xi
K_{\mu\nu}^{(1)} K_{\lambda\sigma}^{(2)} f_1 f_2
= - {1\over 2} f_1 f_2 {(-G)\over (-G_5)^2} \epsilon_{\rho'
\mu\nu\lambda\sigma} \partial_\rho \xi G_5^{\rho\rho'} \tilde{H}^{\mu\nu}
(\tilde{H} G \tilde{H} G\tilde{H})^{\lambda\sigma}. \label{onetwo}
\end{equation}
To simplify this further, we consider the factor
\begin{equation}
\epsilon_{\rho' \mu\nu\lambda\sigma} G_5^{\rho\rho'} \tilde{H}^{\mu\nu}
\tilde{H}^{\lambda\eta}\tilde{H}^{\zeta\sigma},
\end{equation}
which we represent by ($\rho' \mu\nu\lambda\eta\sigma$) to indicate the
sequence in which these six superscripts appear.  Because antisymmetrization of
six 5-valued indices gives zero, one has the relation
\begin{equation}
(\rho'\mu\nu\lambda\eta\sigma) = (\sigma\rho' \mu\nu\lambda\eta) - (\eta
\sigma\rho'\mu\nu\lambda)
+ (\lambda\eta\sigma\rho' \mu\nu) - (\nu\lambda\eta\sigma\rho'\mu) +
(\mu\nu\lambda\eta\sigma\rho'). \label{perms}
\end{equation}
The second term on the right-hand side of eq.~(\ref{perms}) contains
\begin{equation}
\epsilon_{\rho'\mu\nu\lambda\sigma} \tilde{H}^{\sigma\rho'} \tilde{H}^{\mu\nu}
\tilde{H}^{\zeta\lambda}.
\end{equation}
The antisymmetry of $\tilde{H}$ and the fact that the indices are 5-valued
implies that this vanishes.  The last three terms on the right-hand side of (\ref{perms})
are each equal to the negative of the expression on the left.  Thus
\begin{equation}
(\rho' \mu\nu \lambda\eta\sigma) = {1\over 4} (\sigma \rho' \mu\nu
\lambda\eta).
\end{equation}
Substituting this into eq. (\ref{onetwo}) gives
\begin{equation}
- {1\over 8} f_1 f_2 {(-G)\over (-G_5)^2} \epsilon_{\rho'\mu\nu\lambda\sigma}
\partial_\rho \xi G_5^{\rho\sigma} \tilde{H}^{\rho'\mu} \tilde{H}^{\nu\lambda}
(G\tilde{H}G)_{\eta\zeta} \tilde{H}^{\zeta\eta}
= - z_1 f_1 f_2 W.
\end{equation}
The term proportional to $f_2^2$ is evaluated by the same methods and gives
\begin{equation}
\left({1\over 2} z_1^2 - z_2\right) f_2^2 W.
\end{equation}

Assembling the results given above, 
the required general coordinate transformation symmetry is
achieved if, and only if, the function $f$ satisfies the nonlinear partial
differential equation
\begin{equation}
f_1^2 + z_1 f_1 f_2 + \left(z_2 - {1\over 2} z_1^2\right) f_2^2 = 1.
\end{equation}
This is precisely the same equation that was derived in Ref.~\cite{gibbons} as
the condition for self duality of a 4d U(1) gauge theory and in
Ref.~\cite{perry} as the condition for Lorentz invariance of the 6d theory with
$G_{\hat\mu \hat\nu} = \eta_{\hat\mu \hat\nu}$.  As discussed in~\cite{perry},
it has many solutions, but the two solutions of most interest are the ``free'' ($f
= z_1$) solution and the ``Born--Infeld'' solution
\begin{equation}
f = 2 \sqrt{1 + z_1 + {1\over 2} z_1^2 - z_2}.
\end{equation}
The former gives the bosonic part of the 6d supergravity action and the latter
gives the bosonic part of the M theory five-brane action.  In the latter case one
can reexpress the complete bosonic Lagrangian as
\begin{equation}
L = 2 \sqrt{- {\rm det} \Big(G_{\hat\mu \hat\nu} + i G_{\hat\mu\rho} G_{\hat\nu
\lambda} \tilde{H}^{\rho\lambda} / \sqrt{-G_5}\Big)}
+ {1\over 2} \tilde{H}^{\mu\nu} \partial_5 B_{\mu\nu}
- {1\over 4} \epsilon_{\mu\nu\rho\lambda\sigma} {G^{5\rho}\over G^{55}}
\tilde{H}^{\mu\nu} \tilde{H}^{\lambda\sigma} .
\end{equation}
All that remains to construct the M theory five-brane action is to add the
appropriate dependences on fermionic degrees of freedom and to establish local
kappa symmetry.  The result will be reported elsewhere~\cite{aganagic}.

I am grateful to M. Aganagic and C. Popescu for helpful discussions.


\begin{thebibliography}{9}
\bibitem{perry} M. Perry and J.H. Schwarz, ``Interacting Chiral Gauge Fields in
Six Dimensions and Born--Infeld Theory,'' hep-th/9611065.

\bibitem{schwarz} J.H. Schwarz and A. Sen, Nucl. Phys. {\bf B411} (1994) 35,
hep-th/9304154.

\bibitem{gibbons} G.W. Gibbons and D.A. Rasheed, Nucl. Phys. {\bf B454} (1995)
185, hep-th/9506035.

\bibitem{aganagic} M. Aganagic, J. Park, C. Popescu, and J.H. Schwarz, to appear.
\end{thebibliography}
\end{document}